\documentclass[titlepage]{article}

\begin{document}
\title{A proposed theoretical basis for a MOND cosmology: I The inertial frame?}
\author{D F Roscoe \\
Applied Mathematics Department, Sheffield University \\
Sheffield S3 7RH, UK}
\maketitle
\begin{abstract}
This paper is essentially a speculation on the realization of Mach's 
Principle, and we came to the details of the present analysis via the 
formulation of  
two questions: {\it ({\it a}) Can a globally inertial space \& time be
associated with a non-trivial global matter distribution? ({\it b})
If so, what are the general properties of such a global distribution?}

These questions are addressed within the context of an extremely simple
model universe consisting of particles possessing only the property of
enumerability existing in a formless continuum.
Since there are no pre-specified ideas of clocks and rods in this model 
universe, we are forced into 
two fundamental considerations, these being: 
{\it What invariant meanings can be 
given to the concepts of spatial
displacement and elapsed time in this model universe?} \hfill \break 
Briefly, these questions are answered as follows: the spatial displacement of 
a particle 
is defined in terms of its changed relationship with the particle ensemble as a
whole - this is similar to the man walking down a street who can estimate the 
length of his walk by reference to his changed view of the street. 
Once the concept of invariant spatial displacement is established, a
corresponding concept of elapsed time then emerges in a natural way as 
{\lq process'} within the system.

Thus, unlike for example, general relativity, which can be considered as a
theory describing the behaviour of specified clocks and rods in the 
presence of matter, the present analysis can be considered as a rudimentary -
but fundamental - theory of what underlies the concepts of clocks and rods in a
material universe.
In answer to the original two questions, 
this theory tells us that a globally inertial space \& time {\it can} be
associated with a non-trivial global matter distribution, and that this
distribution is {\it necessarily} fractal with $D\,=\,2$.

This latter result is compared with the results of modern surveys of galaxy
distributions which find that such distributions are quasi-fractal with
$D\,\approx \,2$ on the small-to-medium scales, with the situation on the
medium-to-large scales being a topic of considerable debate.
Accordingly, and bearing in mind the extreme simplicity of the model considered,
the observational evidence is consistent with the interpretation that the
analysed point-of-view captures the cosmic reality to a good first-order 
approximation.
We consider the implications of these results.
\hfill \break
\hfill \break
\leftline{\bf Inertia -- Mach -- MOND -- Fractal -- Cosmology}
\end{abstract}
\section{Introduction}
\label{Intro}
The ideas underlying what is now known as {\lq Mach's Principle'} can be
traced to Berkeley (1710, 1721) for which a good contemporary discussion
can be found in Popper (1953).
Berkeley's essential insight, formulated as a rejection of Newton's ideas
of absolute space, was that the motion of any object had no meaning except
insofar as that motion was referred to some other object, or set of objects.
Mach (1960, reprint of 1883 German edition ) went much further than Berkeley 
when he said {\it I have remained to the present day the only one who insists
upon referring the law of inertia to the earth and, in the case of motions
of great spatial and temporal extent, to the fixed stars}.
In this way, Mach formulated the idea that, ultimately, inertial frames should 
be defined with respect to the average rest frame of the visible universe.

It is a matter of history that Einstein was greatly influenced by Mach's 
ideas as expressed in the latter's {\it The Science of Mechanics ...}
(see for example Pais 1982) and believed that they were incorporated in his 
field equations so long as space was closed (Einstein 1950).
The modern general relativistic analysis gives detailed quantitative support to
this latter view,
showing how Mach's Principle can be considered to arise as a consequence of the 
field equations when appropriate conditions are specified on an initial 
hypersurface in a closed evolving universe.
In fact, in answer to Mach's question asking what would happen to inertia
if mass was progressively removed from the universe, Lynden-Bell, Katz \& Bicak
(1995) point out that, in a {\it closed} Friedmann universe the maximum radius 
of this closed universe and the duration of its existence both shrink to zero
as mass is progressively removed.
Thus, it is a matter of record that a satisfactory incorporation of Mach's
Principle within general relativity can be attained when the constraint of
closure is imposed.

However, there is a hardline point of view: in practice, when we talk
of physical space (and the space composed of the set of all inertial frames in 
particular), we mean a space in which 
{\it distances} and {\it displacements} can be determined - but these concepts 
only have any meaning
insofar as they refer to relationships within material systems. Likewise, when
we refer to elapsed physical time, we mean a measurable degree of ordered change
(process) occurring within a given physical system.
Thus, all our concepts of measurable {\lq space \& time'} are irreducibly 
connected to the existence of material systems and to process within such
systems - which is why the closed Friedmann solutions are so attractive.
However, from this, we can also choose to conclude that any theory (for example, 
general relativity notwithstanding its closed Friedmann solutions) that allows 
an internally consistent discussion of an empty inertial
spacetime must be non-fundamental at even the classical level.

To progress, we take the point of view that, since all our concepts of 
measurable {\lq space \& time'} are irreducibly connected to the existence of 
material systems and to process within such systems, then these concepts are, in 
essence, {\it metaphors} for the relationships that exist between the individual 
particles (whatever these might be) within these material systems.
Since the most simple conception of physical space \& time is that provided by
inertial space \& time, we are then led to two simple questions: 
\hfill \break

{\it Is it possible to associate a
globally inertial space \& time with a non-trivial global matter distribution 
and, if it is, what are the fundamental properties of this distribution?}
\hfill \break

In the context of the simple model analysed, 
the present paper finds definitive answers to these questions so that:
\begin{itemize}
\item
{\it 
A globally inertial space \& time {\it can} be associated with a non-trivial
global distribution of matter;
\item
This global distribution is necessarily fractal with $D\,=\,2$.}
\end{itemize}
In the following, we construct a simple model universe, analyse it within the 
context
of the basic questions posed, and consider other significant matters which arise
naturally within the course of the development.
\section{General overview}
\label{Overview}
We start from the position that conceptions of an {\it empty} inertial 
spatio-temporal continuum are essentially non-physical, and are incapable of
providing sound foundations for fundamental theory.
The fact that we have apparently successful theories based exactly 
on such conceptions does not conflict with this statement - so long as we
accept that, in such cases, the empty inertial spatio-temporal continuum is 
understood to be a
metaphor for a deeper reality in which the metric (or inertial) properties of 
this spatio-temporal continuum are somehow projected out of an 
{\it unaccounted-for} universal distribution of material.
For example, according to this view, the fact that general relativity admits an empty 
inertial spatio-temporal continuum as a special case (and was actually originally derived as a generalization
of such a construct) implies that it is based upon such a metaphor - and is 
therefore, according to this view, not sufficiently primitive to act as a basis 
from which fundamental theories of cosmology can be constructed. 

By starting with a model universe consisting
of objects which have no other properties except identity (and hence
enumerability) existing in a formless continuum, we show how it is possible to project
spatio-temporal metric properties from the objects onto the continuum.
By considering idealized dynamical equilibrium conditions (which arise as a limiting 
case of a particular free parameter going to zero), we are then able to show how a
globally inertial spatio-temporal continuum is necessarily identified with a material
distribution which has a fractal dimension $D\,$=$\,2$ in this projected space.
This is a striking result since it bears a very close resemblance
to the cosmic reality for the low-to-medium redshift regime.

However, this idealized limiting case material distribution is distinguished 
from an ordinary material
distribution in the sense that the individual particles of which it is comprised
are each in a state of arbitrarily directed motion, but with equal-magnitude
velocities for all particles - and in this sense is more like a quasi-photon gas
distribution.
For this reason, we interpret the distribution as a rudimentary representation 
of an inertial material vacuum, and present it as the appropriate
physical background within which gravitational processes (as conventionally
understood) can be described as point-source perturbations of an inertial 
spatio-temporal-material background.
We briefly discuss how such processes can arise.
\subsection{Overview of the non-relativistic formalism}
In order to clarify the central arguments and to minimize conceptual problems in this 
initial development, we assume that the model universe is stationary in the sense that 
the overall statistical properties of the material distribution do not evolve in any 
way.
Whilst this was intended merely as a simplifying assumption, it has the fundamental
effect of making the development inherently non-relativistic (in sense that the system 
evolves within a curved metric three-space, rather than being a geodesic structure 
within a spacetime continuum).

The latter consequence arises in the following way: since the model universe is assumed to be
stationary, then there is no requirement to import a pre-determined concept of 
{\lq time'} into the discussion at the beginning - although the qualitative notion of a 
generalized {\lq temporal ordering'} is assumed.
The arguments used then lead to a formal model which allows the natural introduction 
of a generalized temporal ordering parameter, and this formal model is invariant with 
respect to any transformation of this latter parameter which leaves the 
absolute ordering of events unchanged.
This arbitrariness implies that the formal model is incomplete, and 
can only be completed by the imposition of an additional condition which 
constrains the temporal ordering parameter to be identifiable with some model of
physical time.
It is then found that such a model of physical time, defined in terms of
{\lq system process'}, arises automatically from the assumed isotropies within the system.
In summary, the assumption of stationarity leads to the emergent concept of a physical 
{\lq spatio-temporal continuum'} which partitions into a metric three-space together with a 
distinct model of physical time defined in terms of ordered material process in the 
metric three-space.
The fractal $D\,$=$\,2$ inertial universe then arises as an idealized limiting case.
\subsection{Overview of the relativistic formalism}
\label{Relativistic}
The relativistic formalism arises as a natural consequence of relaxing the constraint of a 
{\it stationary} universe.
The formalism is not considered in any detail here but, briefly, its development can be 
described as follows: if the universe is not 
stationary, then it is evolving - and this implies the need for a pre-determined
concept of {\lq time'} to be included in the discussion at the outset.
If this is defined in any of the ways which are, in practice, familiar to us then we can
reasonably refer to it as {\lq local process time'}.
Arguments which exactly parallel those used in the stationary universe case considered
in detail here then lead
to a situation which is identical to that encountered in the Lagrangian formulation
of General Relativity: in that historical case, the equations of motion include a local 
coordinate
time (which corresponds to our local process time) together with a global temporal
ordering parameter, and the equations of motion are invariant with respect to any 
transformation of this
latter parameter which leaves the ordering of {\lq spacetime'} events unchanged.
This implies that the equations of motion are incomplete - and the situation is resolved
there by defining the global temporal ordering parameter to be {\lq particle proper time'}.
The solution we adopt for our evolving universe case is formally identical, so that
everything is described in terms of a metric {\lq spacetime'}.
By considering idealized dynamical equilibrium conditions,
we are led to the concept of an inertial {\lq spacetime'} which is identical to the 
spacetime of special relativity - except that it is now irreducibly associated with a 
fractally distributed {\it relativistic} {\lq photon gas'}.
\section{The starting point}
In \S{\ref{Intro}}, we offered the view that the fundamental significance of 
Mach's Principle arises from its implication of the impossibility of defining 
inertial frames in the absence of material;
or, as a generalization, that it is {\it impossible} to conceive
of a physical spatio-temporal continuum in the absence of material.
It follows from this that, if we are to arrive at a consistent and fundamental
implementation of Mach's Principle, then we need a theory of the world
according to which (roughly speaking) notions of the spatio-temporal continuum are somehow
projected out of primary relationships between objects.
In other words, we require a theory in which notions of metrical space \& time are to be 
considered as metaphors for these primary relationships.
Our starting point is to consider the calibration of a radial measure which
conforms to these ideas.
\hfill \break
\hfill \break
Consider the following perfectly conventional procedure which assumes that
we {\lq know'} what is meant by a given radial displacement, $R$ say.
On a large enough scale ($> 10^8~$light years, say), we can reasonably assume it is
possible to write down a relationship describing the amount of mass contained
within a given spherical volume: say
\begin{equation}
M = U(R),
\label{eqn.0}
\end{equation}
where $U$ is, in principle, determinable.
Of course, a classical description of this type ignores the discrete nature
of real material; however, overlooking this point, such a description is
completely conventional and unremarkable.
Because $M$ obviously increases as $R$ increases, then $U$ is said to be
monotonic, with the consequence that the above relationship can be
inverted to give
\begin{equation}
R\,=\,G(M)
\label{eqn.1}
\end{equation}
which, because (\ref{eqn.0}) is unremarkable, is also unremarkable.
\hfill \break
\hfill \break
In the conventional view, (\ref{eqn.0}) is logically prior to (\ref{eqn.1});
however, it is perfectly possible to reverse the logical priority of
(\ref{eqn.0}) and (\ref{eqn.1}) so that, in effect, we can choose to
{\it define} the radial measure in terms of (\ref{eqn.1}) rather than assume
that it is known by some independent means.
If this is done then, immediately, we have made it impossible to conceive
of radial measure in the absence of material.
With this as a starting point, we are able to construct a completely
Machian Cosmology in a way outlined in the following sections.
\section{A discrete model universe}
\label{sec.3}
The model universe is intended as an idealization of our actual universe, and is defined
as follows:  
\begin{it}
\begin{itemize}
\item it consists of an infinity of identical, {\it but labelled},
discrete material particles which are primitive, possessing no
other properties beyond being countable;
\item {\lq time'} is to be understood, in a qualitative way, as a measure
of {\it process} or {\it ordered change} in the model universe;
\item there is at least one origin about which the distribution of material particles
is statistically isotropic - meaning
that the results of sampling along arbitrary lines of sight over sufficiently long 
characteristic {\lq times'} are independent of the directions of lines of sight;
\item the distribution of material is statistically stationary - 
meaning
that the results of sampling along arbitrary lines of sight over sufficiently long 
characteristic {\lq times'} are independent of sampling epoch.
\end{itemize}
\end{it}
Although concepts of invariant spatio-temporal measurement are implicitly
assumed to exist in this model universe, we make no apriori assumptions about
their quantitative definition, but require that such definitions should arise
naturally from the structure of the model universe and from the following
analysis.
\subsection{The invariant calibration of a radial coordinate
in terms of counting primitive objects.}
\label{subsec.4.1}
At (\ref{eqn.1}), we have already introduced, in a qualitative way, the idea
that the radial magnitude of a given sphere can be {\it defined} in terms of 
the amount of material contained within that sphere and, in this section, we 
seek to make this
idea more rigorous.
To this end, we note that the most primitive invariant that can be conceived
is that based on the counting of objects in a countable set, and we show how this
fundamental idea can be used to define the concept of invariant distance in
the model universe.
\hfill \break
\hfill \break
The isotropy properties assumed for the model universe imply that it is
statistically spherically symmetric about the chosen origin.
If, for the sake of simplicity, it is assumed that the characteristic sampling
times over which the assumed statistical isotropies become exact are
infinitesimal, then the idea of statistical spherical symmetry gives way to
the idea of exact spherical symmetry - thereby allowing the idea of some kind
of rotationally invariant radial coordinate to exist.
As a first step towards defining such an idea, suppose only that the means
exists to define a succession of nested spheres,
$S_1 \subset S_2 \subset ... \subset S_p$, about the chosen origin; since the
model universe with infinitesimal characteristic sampling times is stationary,
then the flux of particles across the spheres is such that
these spheres will always contain fixed numbers of particles, say
$N_1, N_2, ..., N_p$ respectively.
\hfill \break
\hfill \break
Since the only invariant quantity associated with any given sphere, $S$ say,
is the {\it number} of material particles contained within it, $N$ say, then
the only way to associate an invariant radial coordinate, $r$ say, with $S$ is
to {\it define} it according to $r = r_0 f(N)$ where $r_0$ is a fixed
scale-constant having units of {\lq length'}, and the function $f$ is
restricted by the requirements $f(N_a) > f(N_b)$ whenever
$N_a > N_b$, $f(N) > 0$ for all $N>0$, and $f(0)=0$.
To summarize, an invariant calibration of a radial coordinate in the model
universe is given by $r=r_0 f(N)$ where:
\begin{itemize}
\item $f(N_a) > f(N_b)$ whenever $N_a > N_b$;
\item $f(N)>0$ for all $N>0$ and $f(0)=0$.
\end{itemize}
Once a radial coordinate has been invariantly calibrated, it is a matter
of routine to define a rectangular coordinate system based upon this radial
calibration; this is taken as done for the remainder of this paper.
\subsection{The mass model}
At this stage, since no notion of {\lq inertial frame'} has been introduced
then the idea of {\lq inertial mass'} cannot be defined.
However, we have assumed the model universe to be composed of a countable
infinity of labelled - but otherwise indistinguishable - material particles so
that we can associate with each individual particle a property called
{\lq mass'} which quantifies the amount of material in the particle, and is
represented by a scale-constant, $m_0$ say, having units of {\lq mass'}.
\hfill \break
\hfill \break
The radial parameter about any point is {\it defined} by $r=r_0 f(N)$; since
this function is constrained to be monotonic, then its inverse exists
so that, by definition, $N=f^{-1} (r/r_0)$; suppose we now introduce the
scale-constant $m_0$, then $N m_0 = m_0 f^{-1} (r/r_0) \equiv M(r)$ can be
{\it interpreted} as quantifying the total amount of material inside a sphere
of radius $r$ centred on the assumed origin.
Although $r=r_0f(N)$ and $M(r)=N m_0$ are equivalent, the development which
follows is based upon using $M(r)$ as a description of the mass distribution
given as a function of an invariant radial distance parameter, $r$, of
undefined calibration.
\hfill \break
\hfill \break
It is clear from the foregoing discussion that $r$ is defined as a necessarily
discrete parameter.
However, to enable the use of familiar techniques, it will hereafter be supposed
that $r$ represents a continuum - it being understood that a fully consistent
treatment will require the use of discrete mathematics throughout.
\section{The absolute magnitudes of arbitrary 
displacements in the model universe}
\label{sec.4}
We have so far defined, in general terms, an invariant radial
coordinate calibration procedure in terms of the radial distribution of material
valid from the assumed origin, and have noted that such
a procedure allows a routine definition of orthogonal coordinate axes.
Whilst this process has provided a means by which arbitrary displacements can
be described relative to the global material distribution, it does not provide
the means by which an invariant {\it magnitude} can be assigned to such
displacements - that is, there is no metric defined for the model universe.
In the following, we show how the notion of {\lq metric'} can be
considered to be projected from the mass distribution.
\subsection{Change in perspective as an general indicator of displacement in
a material universe}
In order to understand how the notion of {\lq metric'} can be defined, we begin
by noting the following empirical circumstances from our familiar world:
\begin{itemize}
\item
In reality, an observer displaced from one point to another recognizes the fact 
of his own spatial 
displacement by reference to his changed perspective of his (usually local) 
material universe;
\item
the magnitude of this change in perspective provides a measure of the magnitude
of his own spatial displacement.
\end{itemize}
To be more specific, consider an idealized scene consisting of a distributed set 
of many labelled points  
all in a static relationship with respect to each other, plus an observer of
this scene.
Since the labelled points are in a static relationship to each other, then a
subset of them can be used to define a reference frame within which all of the 
other labelled points in the scene occupy fixed positions.
The specification of the observer's directions-of-view onto any two of the 
labelled points in this scene (which are not colinear with him!) uniquely fixes 
the observer's position and hence his {\it perspective} of the whole scene.
Correspondingly, the starting and finishing points of any journey undertaken by the
observer can be specified by the initial and final directions-of-view onto each of 
the two chosen labelled points, and the journey itself can be given an invariant 
description purely in
terms of these initial and final directions-of-view conditions - that is, in
terms of the observer's changed perspective of the whole scene.

To summarize, an observer's perspective of a scene can be considered defined by 
his coordinate position in the defined reference frame plus a direction of view 
onto a specified labelled point within the scene, and an invariant description
of any journey made by the observer of the scene can given in terms of change in
this perspective.
In the following, we show exactly how the concept of {\lq change in
perspective'} can be used to associated invariant magnitudes to coordinate
displacements in the model universe.
\subsection{Perspective in the model universe}
Since, in the present case, we are seeking to give invariant meaning to the
displacement of an arbitrarily chosen particle in the model universe, then we
replace the journeying observer of the foregoing static scene by the chosen 
particle itself.
Additionally, given that the chosen particle lies initially on the constant-mass
surface ($r=constant$) of the mass-model, $M(r)$, then we replace the static 
scene itself by the collection of particles contained within this constant-mass 
surface.

To define perspective information for the chosen particle, we note that there 
is only one distinguished point in the model universe, and that is the origin of 
the mass-model.
Consequently, the most obvious possibility for
perspective information is given by the direction-of-view from the chosen
particle onto the mass-model origin.
Noting how the specification of a constant-mass surface plus the direction to 
the origin uniquely fixes the position of the chosen particle in
the model universe, we conclude that this particle's perspective of the
model universe is completely defined by its constant-mass surface plus its
direction-of-view onto the  mass-model origin.

Finally, we note that, subject to
the magnitude of the normal gradient vector, $|\nabla M |$, being a monotonic
function of $r$, total perspective information is precisely carried by the
normal gradient vector itself.
This follows since the assumed monotonicity of  
$|\nabla M |$ means that this magnitude is in a 1:1 relation with $r$ and so
can be considered to define {\it which} constant-mass surface is observed;
simultaneously, the direction of $\nabla M$ is always radial, and so
defines the direction-of-view from the chosen particle onto the mass-model 
origin.
\hfill \break
\hfill \break
So, to summarize, the perspective of the chosen particle can be
considered defined by the normal gradient vector, ${\bf n} \equiv \nabla M$,
at the particle's position.
\subsection{Change in perspective in the model universe}
We now consider the change in perspective arising from an infinitesimal
change in coordinate position: defining the components of the normal gradient
vector (the perspective) as $n_a \equiv \nabla_a M,~a=1,2,3$, then the
{\it change} in perspective for a coordinate displacement
$d{\bf r} \equiv (dx^1,dx^2,dx^3)$ is given by
\begin{equation}
dn_a = \nabla_j (\nabla_a M) dx^j \equiv g_{ja} dx^j,
~~~ g_{ab} \equiv \nabla_a \nabla_b M,
\label{(1)}
\end{equation}
for which it is assumed that the geometrical connections required to give this
latter expression an unambiguous meaning will be defined in due course.
Given that $g_{ab}$ is non-singular, we now note that (\ref{(1)}) provides a
1:1 relationship between the contravariant vector $dx^a$ (defining change
in the observer's coordinate position) and the covariant vector $dn_a$
(defining the corresponding change in the observer's perspective).
It follows that we can define $dn_a$ as the covariant form of $dx^a$,
so that $g_{ab}$ automatically becomes the mass model metric tensor.
The scalar product $dS^2 \equiv dn_i dx^i$ is then the absolute magnitude of
the coordinate displacement, $dx^a$, defined relative to the change in
perspective arising from the coordinate displacement.
\hfill \break
\hfill \break
The units of $dS^2$ are easily seen to be those of $mass$ only and so, in order
to make them those of $length^2$ - as dimensional consistency requires - we
define the working invariant as $ds^2 \equiv (2 r_0^2/m_0) dS^2 $, where $r_0$
and $m_0$ are scaling constants for the distance and mass scales respectively
and the numerical factor has been introduced for later convenience.
\hfill \break
\hfill \break
Finally, if we want
\begin{equation}
ds^2 \equiv \left({r_0^2 \over 2 m_0}\right) dn_i dx^i \equiv
\left({r_0^2 \over 2 m_0}\right)g_{ij} dx^i dx^j
\label{1a}
\end{equation}
to behave sensibly in the sense that $ds^2=0$ only when $d{\bf r}=0$, then we
must replace the condition of non-singularity of $g_{ab}$ by the condition
that it is strictly positive (or negative) definite;
in the physical context of the present problem, this will be considered to be
a self-evident requirement.
\subsection{The connection coefficients}
We have assumed that the geometrical connection coefficients can be defined
in some sensible way.
To do this, we simply note that, in order to define conservation laws (ie to
do physics) in a Riemannian space, it is necessary to be have a generalized
form of Gausses' divergence theorem in the space.
This is certainly possible when the connections are defined to be the
metrical connections, but it is by no means clear that it is ever possible
otherwise.
Consequently, the connections are assumed to be metrical and so $g_{ab}$,
given at (\ref{(1)}), can be written explicitly as
\begin{equation}
g_{ab} \equiv \nabla_a \nabla_b M
\equiv {\partial^2 M \over \partial x^a \partial x^b}
- \Gamma^k_{ab} {\partial M \over \partial x^k},
\label{(3)}
\end{equation}
where $\Gamma^k_{ab}$ are the Christoffel symbols, and given by
\begin{displaymath}
\Gamma^k_{ab} ~=~ {1 \over 2} g^{kj}
\left( { \partial g_{bj} \over \partial x^{a} }  +
       { \partial g_{ja} \over \partial x^{b} }  -
       { \partial g_{ab} \over \partial x^{j} } \right).
\end{displaymath}
\section{The metric tensor given in terms of the mass model}
\label{sec.6}
It is shown, in appendix \ref{app.A}, how, for an arbitrarily defined mass model,
$M(r)$, (\ref{(3)}) can be exactly resolved to give an explicit form for
$g_{ab}$ in terms of such a general $M(r)$:
Defining
\begin{displaymath}
{\bf r} \equiv (x^1,x^2,x^3),~~
\Phi \equiv {1 \over 2} <{\bf r}| {\bf r}>~~ {\rm and}~~
M' \equiv {d M \over d \Phi }
\end{displaymath}
where $<\cdot|\cdot>$ denotes a scalar product, then it is found that
\begin{equation}
g_{ab} = A \delta_{ab} + B x^i x^j \delta_{ia} \delta_{jb} ,
\label{(4)}
\end{equation}
where
\begin{displaymath}
A \equiv {d_0 M + m_1 \over \Phi},~~~
B \equiv - { A \over 2 \Phi } + { d_0 M' M' \over 2 A \Phi }.
\end{displaymath}
for arbitrary constants $d_0$ and $m_1$ where, as inspection of the
structure of these expressions for $A$ and $B$ shows, $d_0$ is dimensionless
and $m_1$ has dimensions of mass.
Noting that $M$ always occurs in the form $d_0 M + m_1$, it
is convenient to write ${\cal M} \equiv d_0 M + m_1$, and to
write $A$ and $B$ as
\begin{equation}
A \equiv {{\cal M} \over \Phi},~~
B \equiv - \left( {{\cal M} \over 2 \Phi^2} -
{{\cal M}' {\cal M}' \over 2 d_0 {\cal M}} \right).
\label{4a}
\end{equation}
\section{An invariant calibration of the radial scale}
\label{sec.7}
So far, we have assumed an arbitrary calibration for the radial scale; that is,
we have assumed only that $r = f(M)$ where $f$ is an arbitrary monotonic 
increasing function of the mass, $M$.
We seek to find the calibration that incorporates the physical content (that is,
the perspective information) of the metric tensor defined at (\ref{(4)}).
\subsection{The geodesic radial scale}
Using (\ref{(4)}) and (\ref{4a}) in (\ref{1a}), and applying the identities
$x^i dx^j \delta_{ij} \equiv r dr$ and $\Phi \equiv r^2/2$, we find, for an 
arbitrary displacement $d {\bf x}$, the invariant measure:
\begin{displaymath}
ds^2 = \left({r_0^2 \over 2 m_0}\right) \left\{
{ {\cal M} \over \Phi} dx^i dx^j \delta_{ij} -  \Phi 
\left( {{\cal M} \over  \Phi^2} -
{{\cal M}' {\cal M}' \over  d_0 {\cal M}} \right) dr^2 \right\},
\end{displaymath}
which is valid for the arbitrary calibration $r=f(M)$.
If the displacement $d{\bf x}$ is now constrained to be purely radial, then we 
find
\begin{displaymath}
ds^2 =  \left({r_0^2 \over 2 m_0}\right) \left\{
\Phi \left( {{\cal M}' {\cal M}' \over d_0 {\cal M}} \right) dr^2 \right\}.
\end{displaymath}
Use of ${\cal M}' \equiv d{\cal M} / d\Phi$ and $\Phi \equiv r^2/2$
reduces this latter relationship to
\begin{eqnarray}
ds^2 &=& { r_0^2 \over d_0 m_0} \left( d \sqrt{ {\cal M}} \right)^2~~\rightarrow~~
ds = { r_0 \over \sqrt{ d_0 m_0} } d \sqrt{ {\cal M}}~~\rightarrow \nonumber \\
s &=& { r_0 \over \sqrt{ d_0 m_0}}
\left( \sqrt{ {\cal M}} -  \sqrt{ {\cal M}_0} \right), ~~~{\rm where}~~~
{\cal M}_0 \equiv {\cal M}(s=0) \nonumber
\end{eqnarray}
which defines the invariant magnitude of an arbitrary {\it radial}
displacement from the origin purely in terms of the mass-model representation 
${\cal M} \equiv d_0 M + m_1$.
By definition, this $s$ is the radial measure which incorporates the physical 
content of the metric tensor (\ref{(4)}), and so the required calibration is
obtained simply by making the identity $r \equiv s$.
\hfill \break
\hfill \break
To summarize, the natural physical calibration for the radial scale is given by
\begin{equation}
r = { r_0 \over \sqrt{ d_0 m_0}}
\left( \sqrt{ {\cal M}} -  \sqrt{ {\cal M}_0} \right),
\label{4e}
\end{equation}
where ${\cal M}_0$ is the value of ${\cal M}$ at $r=0$.
\subsection{The Euclidean metric}
\label{Euclidean}
Using ${\cal M} \equiv d_0 M + m_1$ and noting that $M(r=0) = 0$ necessarily,
then ${\cal M}_0 = m_1$ and so (\ref{4e}) can be equivalently arranged as
\begin{equation}
{\cal M} = \left[ { \sqrt{d_0 m_0} \over r_0} r +\sqrt{m_1} \right]^2.
\label{4b}
\end{equation}
Using ${\cal M} \equiv d_0 M + m_1$ again, then the mass-distribution function 
can be expressed in terms of the invariant radial displacement as
\begin{equation}
M = m_0 \left( {r \over r_0} \right)^2 +
2 \sqrt{m_0 m_1 \over d_0} \left({ r \over r_0} \right)
\label{4c}
\end{equation}
which, for the particular case $m_1=0$ becomes $M \,=\, m_0 (r /r_0)^2$.
Reference to (\ref{(4)}) shows that, with this mass distribution and $d_0=1$, 
then $g_{ab} = \delta_{ab}$ so that the metric space becomes Euclidean.
Thus, whilst we have yet to show that a globally inertial space can be
associated with a non-trivial global matter distribution (since no temporal
dimension, and hence no dynamics has been introduced), we have shown that a
globally {\it Euclidean} space can be associated with a non-trivial matter
distribution, and that this distribution is necessarily fractal with $D\,=\,2$.
\hfill \break
\hfill \break
Note also that, on a large enough scale and for {\it arbitrary} values of $m_1$, 
(\ref{4c}) shows that radial distance varies as the square-root
of mass from the chosen origin - or, equivalently, the mass varies as $r^2$.
Consequently, on sufficiently large scales Euclidean space is irreducibly 
related to a quasi-fractal, $D\,=\,2$, matter distributions. 
Since $M/r^2\,\approx\, m_0/r_0^2$
on a large enough scale then, for the remainder 
of this paper, the notation $g_0 \equiv m_0/r_0^2$ is employed.
\section{The temporal dimension}
\label{Temporal}
So far, the concept of {\lq time'} has only entered the discussion in the form
of the qualitative definition given in \S\ref{sec.3} - it has not entered in any
quantitative way and, until it does, there can be no discussion of dynamical
processes.

Since, in its most general definition, time is a parameter which orders change 
within a system, then a necessary pre-requisite for its quantitative definition
in the model universe is a notion of change within that universe, and the only
kind of change
which can be defined in such a simple place as the model universe is that of
internal change arising from the spatial displacement of particles.
Furthermore, since the system is populated solely by primitive particles which 
possess only the property of enumerability (and hence quantification in terms of
the {\it amount} of material present) then, in effect, all change is gravitational 
change.
This fact is incorporated into the cosmology to be derived by constraining all
particle displacements to satisfy the Weak Equivalence Principle.
We are then led to a Lagrangian description of particle motions in which the
Lagrange density is degree zero in its temporal-ordering parameter.
From this, it follows that the corresponding Euler-Lagrange equations
form an {\it incomplete} set.
\hfill \break
\hfill \break
The origin of this problem traces back to the fact that, because
the Lagrangian density is degree zero in the temporal ordering parameter,
it is then invariant with respect to any transformation of this parameter
which preserves the ordering.
This implies that, in general, temporal ordering parameters cannot be identified
directly with physical time - they merely share one essential characteristic.
This situation is identical to that encountered in the Lagrangian formulation
of General Relativity; there, the situation is resolved by defining the concept
of {\lq particle proper time'}.
In the present case, this is not an option because the notion of particle
proper time involves the prior definition of a system of observer's clocks -
so that some notion of clock-time is factored into the prior assumptions
upon which General Relativity is built.
\hfill \break
\hfill \break
In the present case, it turns out that the isotropies already imposed on the
system conspire to provide an automatic resolution of the problem which is
consistent with the already assumed interpretation of {\lq time'} as a measure
of ordered change in the model universe.
To be specific, it turns out that the elapsed time associated with any given
particle displacement is proportional, via a scalar field, to the invariant
spatial measure attached to that displacement.
Thus, physical time is defined directly in terms of the invariant measures
of {\it process} within the model universe.
\section{Dynamical constraints in the model universe}
\label{sec.constr}
Firstly, and as already noted, the model universe is populated exclusively by primitive 
particles which possess solely the property of enumeration, and hence quantification.
Consequently, all motions in the model universe are effectively gravitational,
and we model this circumstance by constraining all such motions to satisfy the
Weak Equivalence Principle by which we mean that the trajectory of a body is
independent of its internal constitution.
This constraint can be expressed as:
\noindent
\begin{it}
\begin{quote}
{\bf C1} Particle trajectories are independent of the specific mass values
of the particles concerned;
\end{quote}
\end{it}
\indent
Secondly, given the isotropy conditions imposed on the model universe from the 
chosen origin, symmetry arguments lead to the conclusion that the net action of the whole
universe of particles acting on any given single particle is such that any
net acceleration of the particle must always appear to be directed
through the coordinate origin.
Note that this conclusion is {\it independent} of any notions of retarded
or instantaneous action.
This constraint can then be stated as:
\noindent
\begin{it}
\begin{quote}
{\bf C2} Any acceleration of any given material particle must necessarily be along 
the line connecting the particular particle to the coordinate origin.
\end{quote}
\end{it}
\indent
\section{Gravitational trajectories}
Suppose $p$ and $q$ are two arbitrarily chosen point coordinates on the
trajectory of the chosen particle, and suppose that (\ref{1a}) is
integrated between these points to give the scalar invariant
\begin{equation}
I(p,q) = \int^q_p \left({1 \over \sqrt{2 g_0}}\right) \sqrt{dn_i dx^i}
\equiv
\int^q_p \left({1 \over \sqrt{2 g_0}}\right)\sqrt{ g_{ij} dx^i dx^j }.
\label{(2)}
\end{equation}
Then, in accordance with the foregoing interpretation, $I(p,q)$ gives a scalar
record of how the particle has moved between $p$ and $q$ defined with respect
to the particle's continually changing relationship with the mass model,
$M(r)$.

Now suppose $I(p,q)$ is minimized with respect to choice of the trajectory
connecting $p$ and $q$, then this minimizing trajectory can be interpreted
as a geodesic in the Riemannian space which has $g_{ab}$ as its metric tensor.
Given that $g_{ab}$ is defined in terms of the mass model $M(r)$ - the
existence of which is independent of any notion of {\lq inertial mass'},
then the existence of the metric space, and of geodesic curves within it, is
likewise explicitly independent of any concept of inertial-mass.
It follows that the identification of the particle trajectory ${\bf r}$ with
these geodesics means that particle trajectories are similarly independent
of any concept of inertial mass, and can be considered as the modelling step
defining that general subclass of trajectories which conform to that
characteristic phenomenology of gravitation defined by condition
{\bf C1} of \S\ref{sec.constr}.
\section{The equations of motion}
\label{sec.7a}
Whilst the mass distribution, represented by ${\cal M}$, has been explicitly
determined in terms of the geodesic distance at (\ref{4b}), it is convenient to
develop the theory in terms of unspecified ${\cal M}$.
\hfill \break
\hfill \break
The geodesic equations in the space with the metric tensor (\ref{(4)})
can be obtained, in the usual way, by defining the Lagrangian density
\begin{equation}
{\cal L} \equiv \left({1 \over \sqrt{2 g_0}}\right)
\sqrt{g_{ij} {\dot x}^i {\dot x}^j}
= \left({1 \over \sqrt{2 g_0}}\right)
\left( A <{\dot {\bf r}}|{\dot {\bf r}}> + B {\dot \Phi}^2 \right)^{1/2},
\label{4d}
\end{equation}
where ${\dot x^i} \equiv dx^i/dt$, etc.,
and writing down the Euler-Lagrange equations
\begin{eqnarray}
2 A {\ddot {\bf r}} &+&
\left(2A' {\dot \Phi} - 2{{\dot {\cal L}} \over {\cal L}} A \right)
{\dot {\bf r}} + \left(  B' {\dot \Phi}^2 + 2B {\ddot \Phi} -
A' <{\dot {\bf r}}|{\dot {\bf r}}>
- 2{{\dot {\cal L}} \over {\cal L}} B {\dot \Phi} \right) {\bf r} 
\nonumber \\
&=& 0,
\label{(5)}
\end{eqnarray}
where ${\dot {\bf r}} \equiv d {\bf r}/dt$ and $A' \equiv dA/d \Phi$, etc.
By identifying particle trajectories with geodesic curves, this equation is
now interpreted as the equation of motion, referred to the chosen origin,
of a single particle satisfying condition {\bf C1} of
\S\ref{sec.constr}.
\hfill \break
\hfill \break
However, noting that the variational principle, (\ref{(2)}), is of order
zero in its temporal ordering parameter, we can conclude that the principle is
invariant with respect to arbitrary transformations of this parameter; in turn, this means
that the temporal ordering parameter cannot be identified with physical time.
This problem manifests itself formally in the statement that
the equations of motion (\ref{(5)}) do not form a complete set, so that it becomes
necessary to specify some extra condition to close the system.

A similar circumstance arises in General Relativity theory when the
equations of motion are derived from an action integral which is formally
identical to (\ref{(2)}).
In that case, the system is closed by specifying
the arbitrary time parameter to be the {\lq proper time'}, so that
\begin{equation}
d \tau = {\cal L}(x^j, dx^j)~~\rightarrow~~
{\cal L}(x^j, {d x^j \over d \tau}) = 1,
\label{(5a)}
\end{equation}
which is then considered as the necessary extra condition required to close
the system.
In the present circumstance, we are rescued by the, as yet, unused condition
{\bf C2}.
\section{Physical time}
\label{sec.9}
\subsection{Completion of equations of motion}
Consider {\bf C2}, which states that any particle accelerations must necessarily be
directed through the coordinate origin.
This latter condition simply means that the
equations of motion must have the general structure
\begin{displaymath}
{\ddot {\bf r}} = G(t,{\bf r},{\dot {\bf r}}) {\bf r},
\end{displaymath}
for scalar function $G(t,{\bf r},{\dot {\bf r}})$.
In other words, (\ref{(5)}) satisfies condition {\bf C2} if the coefficient of
${\dot {\bf r}}$ is zero, so that
\begin{equation}
\left(2A' {\dot \Phi} - 2{{\dot {\cal L}} \over {\cal L}} A \right) = 0
~~~\rightarrow
{A'\over A} {\dot \Phi} = {{\dot {\cal L}} \over {\cal L}} ~~\rightarrow~~
{\cal L} = k_0 A,
\label{(6)}
\end{equation}
for arbitrary constant $k_0$ which is necessarily positive since
$A>0$ and ${\cal L}>0$.
The condition (\ref{(6)}), which guarantees ({\bf C2}), can be considered as
the condition required to close the incomplete set (\ref{(5)}), and is directly
analogous to (\ref{(5a)}), the condition which defines {\lq proper time'} in General 
Relativity.
\subsection{Physical time defined as process}
\label{sec.9a}
Equation (\ref{(6)}) can be considered as that equation which removes the
pre-existing arbitrariness in the {\lq time'} parameter by {\it defining}
physical time:-
from (\ref{(6)}) and (\ref{4d}) we have
\begin{eqnarray}
{\cal L}^2 &=& k_0^2 A^2 ~ \rightarrow ~
A <{\dot {\bf r}}|{\dot {\bf r}}> + B {\dot \Phi}^2 = 2 g_0 k_0^2 A^2
~ \rightarrow ~ \nonumber \\
g_{ij} {\dot x}^i {\dot x}^j &=& 2 g_0 k_0^2 A^2
\label{(7)}
\end{eqnarray}
so that, in explicit terms, physical time is {\it defined} by the
relation
\begin{equation}
dt^2 = \left( {1 \over 2 g_0 k_0^2 A^2} \right) g_{ij} dx^i dx^j,~~~
{\rm where}~~A \equiv {{\cal M} \over \Phi}.
\label{(8)}
\end{equation}
In short, the elapsing of time is given a direct physical interpretation in
terms of the process of {\it displacement} in the model universe.

Finally, noting that, by (\ref{(8)}), the dimensions of $k_0^2$ are those of
$L^6 /[T^2 \times M^2]$, then the fact that $g_0 \equiv m_0/r_0^2$ 
(cf \S\ref{sec.7}) suggests the change of notation $k_0^2 \propto v_0^2 /g_0^2$, where 
$v_0$ is a constant having the dimensions (but not the interpretation) of {\lq velocity'}.
So, as a means of making the dimensions which appear in the development more
transparent, it is found convenient to use the particular replacement
$k_0^2 \equiv v_0^2 /(4 d_0^2  g_0^2)$, where $d_0$ is the dimensionless
global constant introduced in \S\ref{sec.6}.
With this replacement, the {\it definition} of physical time, given at
(\ref{(8)}), becomes
\begin{equation}
dt^2 = \left( {4 d_0^2 g_0 \over v_0^2 A^2} \right) g_{ij} dx^i dx^j.
\label{(8a)}
\end{equation}
Since, as is easily seen from the definition of $g_{ab}$ given in \S\ref{sec.6},
$g_{ij} dx^i dx^j$ is necessarily finite and non-zero for a non-trivial
displacement $d{\bf r}$
\subsection{The necessity of $v_0^2 \, \neq \,0$}
\label{sec.12.3}
Equation (\ref{(8a)}) provides a definition of physical time in terms of basic
process (displacement) in the model universe.
Since the parameter $v_0^2$ occurs no where else, except in its explicit position
in (\ref{(8a)}), then it is clear that setting $v_0^2 = 0$ is equivalent to
physical time becoming undefined.
Therefore, of necessity, $v_0^2 \neq 0$ and all non-zero finite displacements
are associated with a non-zero finite elapsed physical time.
\section{The cosmological potential}
The model is most conveniently interpreted when expressed in potential
terms and so, in the following, it is shown how this is done.
\subsection{The equations of motion: potential form}
\label{sec.10}
From \S\ref{sec.9}, when (\ref{(6)}) is used in (\ref{(5)}) there results
\begin{equation}
2A {\ddot {\bf r}} +
\left( B' {\dot \Phi}^2 + 2B {\ddot \Phi} - A' <{\dot {\bf r}}|{\dot {\bf r}}>
- 2{A' \over A} B {\dot \Phi}^2 \right) {\bf r} = 0.
\label{(9)}
\end{equation}
Suppose we define a function $V$ according to
$V \equiv C_0  -<{\dot {\bf r}}|{\dot {\bf r}}>/2$, for some arbitrary
constant $C_0$; then, by (\ref{(7)})
\begin{equation}
V \equiv C_0 -{1 \over 2}<{\dot {\bf r}}|{\dot {\bf r}}> =
C_0 -{v_0^2 \over 4 d_0^2\, g_0} A + {B \over 2A}{\dot \Phi}^2,
\label{(10)}
\end{equation}
where $A$ and $B$ are defined at (\ref{4a}).
With unit vector, ${\hat {\bf r}}$, then appendix {\ref{app.C} shows how
this function can be used to express (\ref{(9)}) in the potential form
\begin{equation}
{\ddot {\bf r}} = -{d V \over d r} {\hat {\bf r}}
\label{(11)}
\end{equation}
so that $V$ is a potential function, and $C_0$ is the arbitrary constant usually
associated with a potential function.
\subsection{The potential function, $V$, as a function of $r$}
\label{sec.centres}
From (\ref{(10)}), we have
\begin{displaymath}
2 C_0\,-\, 2 V \,=\,
{\dot r}^2 + r^2 {\dot \theta}^2 = {v_0^2 \over 2 d_0^2\,g_0 } A -
{B \over A} r^2 {\dot r}^2
\end{displaymath}
so that $V$ is effectively given in terms of $r$ and ${\dot r}$.
In order to clarify things further, we now eliminate the explicit appearance of
${\dot r}$.
Since all forces are central, then angular momentum is conserved; consequently,
after using conserved angular momentum, $h$, and the definitions
of $A$, $B$ and ${\cal M}$ given in \S\ref{sec.6}, the foregoing 
equations can be written as
\begin{eqnarray}
2C_0 &-& 2 V ~~\,=\, \nonumber \\
{\dot r}^2 &+& r^2 {\dot \theta}^2\,=\,  v_0^2 +
{4 v_0^2 \over r} \sqrt{{ m_1 \over d_0 g_0}}
+ {d_0 -1\over r^2} \left( {6 m_1 v_0^2 \over d_0^2\, g_0} -  h^2 \right)
\nonumber \\
&+& { 2 \over r^3 } \sqrt{ {d_0 m_1 \over g_0} }
\left( {2 m_1 v_0^2 \over d_0^2\, g_0} - h^2 \right)
+ {1 \over r^4} { m_1 \over g_0}
\left( { m_1 v_0^2 \over d_0^2\, g_0}  - h^2 \right)
\label{11e}
\end{eqnarray}
so that $V(r)$ is effectively given by the right-hand side of (\ref{11e}).
\section{A discussion of the potential function}
It is clear from (\ref{11e}) that $m_1$ plays the role of the mass of the
central source which generates the potential, $V$.
A relatively detailed description of the behaviour of $V$ is given in appendix 
\ref{app.B}, where we find that there are two distinct classes of solution
depending on the free parameters of the system. 
These classes can be described as:
\begin{itemize}
\item A constant potential universe within
which all points are dynamically indistinguishable;
this corresponds to an inertial material universe, and
arises in the case $m_1=0,~d_0=1$;
\item All other possibilities give rise to a {\lq distinguished origin'} 
universe in which either: 
\begin{itemize}
\item there is a singularity at the centre, $r = 0$;
\item or there is no singularity at $r = 0$ and, instead, the origin is
the centre of a non-trivial sphere of radius $R_{min}  > 0$
which acts as an impervious boundary between the exterior universe and
the potential source. 
In effect, this sphere provides the source with a non-trivial spatial
extension so that the classical notion of the massive point-source is
avoided.
\end{itemize}
\end{itemize}
Of these possibilities, the constant potential universe is the one which 
provides positive answers to our originally posed questions, and it is this 
which is discussed in detail in the following sections.

However, of the two cases in the distinguished origin universe, the no-singularity
case offers the interesting possibility of being able to model the gravitational
effects created by a central massive source, but without the non-physical 
singularity at the origin. 
This case is mentioned here for future reference.
\section{The fractal $D\,$=$\,2$ inertial universe}
\label{Fractal}
Reference to (\ref{11e}) shows that the parameter choice $m_1=0$ and $d_0=1$
makes the potential function constant everywhere, whilst (\ref{4c}) shows how, 
for this case,
universal matter in an equilibrium universe is necessarily distributed as an 
exact fractal with $D=2$.
Thus, the fractal $D =2$ material universe is necessarily a globally
inertial equilibrium universe, and the questions originally posed in 
\S{\ref{Intro}} are finally answered.
\subsection{Implications for theories of gravitation}
Given that gravitational phenomena are usually considered to arise
as mass-driven perturbations of flat inertial backgrounds, then the foregoing
result - to the effect that the inertial background is necessarily associated
with a non-trivial fractal matter distribution - must necessarily give rise to
completely new perspectives about the nature and properties of gravitational
phenomena.
However, as we show in \S\ref{sec.15.1}, the kinematics in this inertial universe is unusual, and 
suggests that the inertial material distribution is more properly interpreted 
as a quasi-photon fractal gas out of which (presumably) we can consider ordinary 
material to condense in some fashion.
\subsection{The quasi-photon fractal gas}
\label{sec.15.1}
For the case $m_1 = 0$, $d_0 = 1$, the definition $M$ at (\ref{4c}) together
with the definitions of $A$ and $B$ in \S\ref{sec.6} give 
\begin{displaymath}
A \,=\,{2 m_0 \over r_0^2},~~~B\,=\,0
\end{displaymath}
so that, by (\ref{(10)}) (remembering that $g_0 \equiv   m_0/r_0^2$) we have
\begin{equation}
<{\dot {\bf r}}|{\dot {\bf r}}> \,=\,v_0^2
\label{(12)}
\end{equation}
for all displacements in the model universe.
It is (almost) natural to assume that the constant $v_0^2$ in (\ref{(12)}) simply refers 
to the constant velocity of any given particle, and likewise to assume
that this can differ between particles.
However, each of these assumptions would be wrong since - as we now show - $v_0^2$ is,
firstly, more properly interpreted as a conversion factor from spatial to temporal 
units and, secondly, is a {\it global} constant which applies equally to all particles.

To understand these points, we begin by noting that (\ref{(12)}) is a special case of 
(\ref{(7)}) and so, by (\ref{(8)}), is more accurately written as
\begin{equation}
dt^2 \,=\, {1 \over v_0^2} < d{\bf r}| d{\bf r}>
\label{(13)}
\end{equation}
which, by the considerations of \S\ref{sec.9a}, we recognize as the 
{\it definition}
of the elasped time experienced by any particle undergoing a spatial displacement 
$d{\bf r}$ in the model inertial universe.
Since this universe is isotropic about all points, then there is nothing which can
distinguish between two separated particles (other than their separateness) undergoing
displacements of equal magnitudes; consequently, each must be considered to have
experienced equal elapsed times.
It follows from this that $v_0^2$ is not to be considered as a locally defined
particle velocity, but is a {\it globally} defined constant which has the effect
of converting between spatial and temporal units of measurement.

We now see that the model inertial universe, with (\ref{(13)}) as a global relationship,
bears a close formal resemblance to a universe filled purely with Einsteinien photons -
the difference being, of course, that the particles in the model inertial universe are 
assumed to be countable and to have mass properties.
This formal resemblance means that the model inertial universe can be likened to a 
quasi-photon fractal gas universe.
\section{A quasi-fractal mass distribution law, 
$M \approx r^2$: the evidence}
\label{sec.evidence}
A basic assumption of the {\it Standard Model} of modern cosmology is that, on 
some scale, the
universe is homogeneous; however, in early responses to suspicions that the
accruing data was more consistent with Charlier's conceptions of an hierarchical
universe (Charlier, 1908, 1922, 1924) than with the requirements of the
{\it Standard Model}, de Vaucouleurs (1970) showed that, within wide limits,
the available data satisfied a mass distribution law $M \approx r^{1.3}$,
whilst Peebles (1980) found $M \approx r^{1.23}$.
The situation, from the point of view of the {\it Standard Model}, has continued
to deteriorate with the growth of the data-base to the point that,
(Baryshev et al (1995))
\begin{it}
\begin{quote}
...the scale of the largest inhomogeneities ({\rm discovered to date}) is
comparable with the extent of the surveys, so that the largest known structures
are limited by the boundaries of the survey in which they are detected.
\end{quote}
\end{it}
For example, several recent redshift surveys, such as those performed by
Huchra et al (1983), Giovanelli and Haynes (1986), De Lapparent et al (1988),
Broadhurst et al (1990), Da Costa et al (1994) and Vettolani et al (1994) etc
have discovered massive structures such as sheets, filaments, superclusters and
voids, and show that large structures are common features of the observable
universe; the most significant conclusion to be drawn from all of these
surveys is that the scale of the largest inhomogeneities observed is comparable
with the spatial extent of the surveys themselves.

In recent years, several quantitative analyses of both pencil-beam and wide-angle 
surveys of galaxy distributions have been performed: three recent examples are
give by Joyce, Montuori \& Labini (1999) who analysed the CfA2-South catalogue 
to 
find fractal behaviour with $D\,$=$\,1.9 \pm 0.1$; Labini \& Montuori (1998) analysed
the APM-Stromlo survey to find fractal behaviour with $D\,$=$\,2.1 \pm 0.1$, 
whilst 
Labini, Montuori \& Pietronero (1998) analysed the Perseus-Pisces survey to 
find fractal behaviour with $D\,$=$\,2.0 \pm 0.1$.
There are many other papers of this nature in the literature all supporting the
view that, out to medium depth at least, galaxy distributions appear to be
fractal with $D\,$$\approx$$\,2$.

This latter view is now widely accepted (for example, see Wu, Lahav \& Rees 
(1999)), and the 
open question has become whether or not there is a transition to homogeneity on some 
sufficiently large scale.
For example, Scaramella et al (1998) analyse the ESO Slice Project redshift
survey, whilst Martinez et al (1998) analyse the Perseus-Pisces, the APM-Stromlo 
and the 1.2-Jy IRAS redshift surveys, with both groups finding evidence for a 
cross-over to homogeneity at large scales.
In response, the Scaramella et al analysis has been criticized on various grounds 
by Joyce et al (1999).

So, to date, evidence that galaxy distributions are fractal with $D \approx 2$ 
on small to medium scales is widely accepted, but there is a lively open debate
over the existence, or otherwise, of a cross-over to homogeneity on large scales.

To summarize, there is considerable debate centered around the question
of whether or not the material in the universe is distributed fractally
or not, with supporters of the big-bang picture arguing that, basically, it
is not, whilst the supporters of the fractal picture argue that it is
with the weight of evidence supporting $D \approx 2$.
This latter position corresponds exactly with the picture predicted by
the present approach.
\section{Summary and Conclusions}
Prompted by the questions 
\hfill \break
\hfill \break
{\it Is it possible to associate a
globally inertial space \& time with a non-trivial global matter distribution 
and, if it is, what are the fundamental properties of this distribution?}
\hfill\break
\hfill \break
we have analysed a very simple model universe, consisting solely of an infinite
ensemble of particles, possessing only the property of enumerability, existing
in a formless continuum and with the ensemble being in a statistically 
stationary state.
No concepts of rods or clocks were imported into this system, and we required
that invariant meanings for spatial and temporal intervals should arise from
within the ensemble itself.

The notion of the spatial displacement of a particle was given meaning using our 
common experience - according to which we recognize our own spatial displacements,
and their magnitudes, by making reference to our changed views of our local
environment and the magnitudes of such changes - and not by referral to formal
measuring rods.
The formal modelling of this experience led, in \S\ref{Euclidean}, to the
conclusions that, within the model universe:
\begin{itemize}
\item
On sufficiently large scales, space is necessarily Euclidean (to any
required degree of approximation) and is
irreducibly associated with a quasi-fractal, $D=2$, distribution of material
within the model universe.
\item
In the ideal limiting case of a particular parameter going to zero, space is
necessarily
identically Euclidean and is irreducibly related to a fractal, $D=2$,
distribution of material within the model universe.
\end{itemize}
This procedure then led, via symmetry arguments, to a formal definition of 
{\lq elapsed time'} within the model universe as an invariant measure of
{\it ordered process} within that universe. It is to be noted that this 
is in accord with the way in which we actually experience the passage of time
in our lives - as the accumulation of ordered process, and not by continual 
reference to formal cyclic clocks. 

With these definitions of invariant spatial displacement and invariant elapsed 
time in place, we were then able to answer the original two questions within 
the context of the model universe so that, finally, we could say: 
\begin{itemize} 
\item 
On sufficiently large scales,  
space \& time is necessarily inertial (to any required degree of approximation),
and is irreducibly associated with a quasi-fractal, $D=2$, distribution of 
material within the model universe;
\item
In the ideal limiting case of a particular parameter going to zero, a
globally inertial space \& time is irreducibly related to a fractal,
$D=2$, distribution of material within the model universe.
\end{itemize}
However, the latter ideal inertial universe is distinguished in the sense that
whilst all the particles within it have arbitrarily directed motions, the
particle velocities all have {\it equal} magnitude.
In this sense, the globally inertial model universe is more accurately to be 
considered as a quasi-photon gas universe than the universe of our 
macroscopic experience.
In other words, it looks more like a crude model of a material vacuum
than the universe of our direct experience. 

This result is to be compared with the distribution of galaxies in our directly
observable universe which approximates very closely perfectly inertial 
conditions, and which appears to be fractal with $D\,$$\approx$$\,2$ on the 
small-to-medium scale at least.
If we make the simple assumption that the distribution of ponderable matter
traces the distribution of the material vacuum then,
given the extreme simplicity of the analysed model, this latter correspondence
between between the model's statements and the cosmic reality lends strong support to
the idea that our intuitively experienced perceptions of physical space and time
are projected out of relationships, and changing relationships, between the
particles (whatever these might be) in the material universe in very much the 
way described.

The foregoing considerations have fundamental consequences for
gravitation theory: specifically, since gravitational phenomena are 
conventionally considered to arise as mass-driven perturbations of a flat 
inertial background, then the phenomonology predicted by the analysis - that a
flat inertial background is irreducibly associated with a non-trivial fractal 
distribution of material - must necessarily lead to novel insights into the 
nature and causes of gravitational phenomena.

Finally, as we have noted, the restriction that the ensemble should be 
statistically stationary (imposed initially for simplicity) was equivalent to 
making the analysis non-relativistic.
The relativistic counterpart of the foregoing analysis arises from a
consideration of a non-stationary universe, and gives rise to a model universe
in which the flat spacetime of special relativity is irreducible associated
with a relativistically invariant material vacuum of fractal dimension.
\appendix{}
\section{A Resolution of the Metric Tensor}
\label{app.A}
The general system is given by
\begin{displaymath}
g_{ab} = {\partial^2 M \over \partial x^a \partial x^b}
-\Gamma^k_{ab} {\partial M \over \partial x^k},
\end{displaymath}
\begin{displaymath}
\Gamma^k_{ab} ~\equiv~ {1 \over 2} g^{kj}
\left( { \partial g_{bj} \over \partial x^{a} }  +
       { \partial g_{ja} \over \partial x^{b} }  -
       { \partial g_{ab} \over \partial x^{j} } \right),
\end{displaymath}
and the first major problem is to express $g_{ab}$ in terms of the
reference scalar, $M$.
The key to this is to note the relationship
\begin{displaymath}
{\partial^2 M \over \partial x^a \partial x^b} =
M' \delta_{ab} + M'' x^a x^b ,
\end{displaymath}
where $M' \equiv dM/d \Phi$, $M'' \equiv d^2M/d \Phi^2$ and
$\Phi \equiv <{\bf r}|{\bf r}>/2$, since this immediately suggests the general
structure
\begin{equation}
g_{ab} = A \delta_{ab} + B x^a x^b,
\label{A1}
\end{equation}
for unknown functions, $A$ and $B$.
It is easily found that
\begin{displaymath}
g^{ab} = {1 \over A } \left[ \delta_{ab} - \left( { B \over A+2B \Phi} \right)
x^a x^b \right]
\end{displaymath}
so that, with some effort,
\begin{displaymath}
\Gamma^k_{ab} = {1 \over 2A} H_1 - \left( { B \over 2A(A+2B \Phi)} \right) H_2
\end{displaymath}
where
\begin{eqnarray}
H_1 & = & A' (x^a \delta_{bk} + x^b \delta_{ak} - x^k \delta_{ab}) \nonumber \\
& + & B' x^a x^b x^k + 2B \delta_{ab} x^k \nonumber
\end{eqnarray}
and
\begin{eqnarray}
H_2 & = & A' (2 x^a x^b x^k - 2 \Phi x^k \delta_{ab} )  \nonumber \\
& + & 2 \Phi B' x^a x^b x^k + 4 \Phi B x^k \delta_{ab}. \nonumber
\end{eqnarray}
Consequently,
\begin{eqnarray}
g_{ab} &=& {\partial^2 M \over \partial x^a \partial x^b}
-\Gamma^k_{ab} {\partial M \over \partial x^k}
\equiv  \delta_{ab} M' \left( {A + A' \Phi \over A+2B \Phi} \right)
\nonumber \\
&+& x^a x^b \left( M'' - M' \left( {A' + B' \Phi \over A+2B \Phi } \right) \right).
\nonumber
\end{eqnarray}
Comparison with (\ref{A1}) now leads directly to
\begin{eqnarray}
A &=& M' \left( {A + A' \Phi \over A+2B \Phi} \right) =
M' \left( (A \Phi)' \over A+2B \Phi \right),   \nonumber \\
B &=&  M'' - M' \left( {A' + B' \Phi \over A+2B \Phi } \right). \nonumber
\end{eqnarray}
The first of these can be rearranged as
\begin{displaymath}
B={M' \over 2 \Phi} \left( (A \Phi)' \over A \right) - {A \over 2 \Phi}
\end{displaymath}
or as
\begin{displaymath}
\left(M' \over A+2 B \Phi \right) = {A \over (A \Phi)' },
\end{displaymath}
and these expressions can be used to eliminate $B$ in the second equation.
After some minor rearrangement, the resulting equation is easily integrated
to give, finally,
\begin{displaymath}
A \equiv {d_0 M + m_1 \over \Phi},~~~
B \equiv - { A \over 2 \Phi } + { d_0 M' M' \over 2 A \Phi }.
\end{displaymath}
\section{Conservative Form of Equations of Motion}
\label{app.C}
From (\ref{(10)}), we have
\begin{equation}
V \equiv -{1 \over 2}<{\dot {\bf r}}|{\dot {\bf r}}> =
-{k_0^2 A \over 2} + {B \over 2A}{\dot \Phi}^2,
\label{B1}
\end{equation}
from which we easily find
\begin{displaymath}
{d V \over dr}  \equiv {\partial V \over \partial r} +
{\partial V \over \partial {\dot r}} { {\ddot r} \over {\dot r}}
\end{displaymath}
\begin{displaymath}
= {-k_0^2 A' \over 2} r +
{ {\dot \Phi}^2 r \over 2A} \left( B'- {A' B \over A} \right)
+{B \over A} \left( r {\dot r}^2 + r^2 {\ddot r} \right).
\end{displaymath}
Since ${\dot r}^2 + r {\ddot r} = {\ddot \Phi}$, then the above expression
leads to
\begin{displaymath}
{d V \over d r}{\bf {\hat r}} =
\left( {-k_0^2 A' \over 2} +
{B' \over 2A} {\dot \Phi}^2  -{A' B \over 2 A^2} {\dot \Phi}^2
+{B \over A}  {\ddot \Phi} \right) {\bf r}.
\end{displaymath}
Writing (\ref{(11)}) as
\begin{displaymath}
2A {\bf {\ddot r}} + 2A {d V \over d r} {\bf {\hat r}} = 0,
\end{displaymath}
and using the above expression, we get the equation of motion as
\begin{equation}
2A {\bf {\ddot r}} +
\left( -k_0^2 A A'  +
B' {\dot \Phi}^2  -{A' B \over A} {\dot \Phi}^2
+ 2B {\ddot \Phi} \right) {\bf r} = 0. 
\label{B2}
\end{equation}
Finally, from (\ref{B1}), we have
\begin{displaymath}
k_0^2 A = {B \over A}{\dot \Phi}^2 + <{\dot {\bf r}}|{\dot {\bf r}}>,
\end{displaymath}
which, when substituted into (\ref{B2}), gives (\ref{(9)}).
\section{Outline analysis of the potential function}
\label{app.B}
It is quite plain from (\ref{11e}) that, for any $m_1 \neq 0$, then the model
universe has a preferred centre and that the parameter $m_1$ (which has dimensions of 
mass) plays a role in the potential $V$ which is analogous to the source mass in a
Newtonian spherical potential - that is, the parameter $m_1$ can be identified as
the mass of the potential source in the model universe.
However, setting $m_1 = 0$ is not sufficient to guarantee a constant potential
field since any $d_0 \neq 1$ also provides the model universe with a preferred centre.
The role of $d_0$ is most simply discussed in the limiting case of $m_1 = 0$: in
this case, the second equation of (\ref{11e}) becomes
\begin{equation}
{\dot r}^2 + r^2 {\dot \theta}^2\,=\,  v_0^2 
- (d_0 -1){ h^2 \over r^2}.
\label{11f}
\end{equation}
If $d_0 < 1$ then $|{\bf {\dot r}}| \rightarrow \infty$ as $r \rightarrow 0$ so that
a singularity exists.
Conversely, remembering that $v_0^2 > 0$ (cf \S\ref{sec.12.3}) 
then, if $d_0 > 1$,
equation (\ref{11f}) restricts real events to the exterior of the sphere
defined by $r^2 = (d_0-1)h^2/v_0^2 $.
In this case, the singularity is avoided and the central {\lq massless particle'} is 
given the physical property of {\lq finite extension'}.
In the more realistic case for which $m_1 > 0$, reference to (\ref{11e}) shows that
the $r=0$ singularity is completely avoided whenever $h^2  >  m_1 v_0^2/d_0^2 g_0$ 
since then a {\lq finite extension'} property for the central massive particle 
always exists.
Conversely, a singularity will necessarily exist whenever $h^2  \leq  m_1 v_0^2/d_0^2 g_0$.

In other words, the model universe has a preferred centre when either $m_1 > 0$, in which
case the source of the potential is a massive central particle having various properties
depending on the value of $d_0$, or when $m_1 = 0$ and $d_1 \neq 0$.

\label{lastpage}
\end{document}